\documentclass{ieeeaccess}
\usepackage{cite}
\usepackage{amsmath,amssymb,amsfonts}
\usepackage{algorithmic}
\usepackage{graphicx}
\usepackage{textcomp}

\usepackage{bm}
\makeatletter
\AtBeginDocument{\DeclareMathVersion{bold}
\SetSymbolFont{operators}{bold}{T1}{times}{b}{n}
\SetSymbolFont{NewLetters}{bold}{T1}{times}{b}{it}
\SetMathAlphabet{\mathrm}{bold}{T1}{times}{b}{n}
\SetMathAlphabet{\mathit}{bold}{T1}{times}{b}{it}
\SetMathAlphabet{\mathbf}{bold}{T1}{times}{b}{n}
\SetMathAlphabet{\mathtt}{bold}{OT1}{pcr}{b}{n}
\SetSymbolFont{symbols}{bold}{OMS}{cmsy}{b}{n}
\renewcommand\boldmath{\@nomath\boldmath\mathversion{bold}}}
\makeatother

\def\BibTeX{{\rm B\kern-.05em{\sc i\kern-.025em b}\kern-.08em
    T\kern-.1667em\lower.7ex\hbox{E}\kern-.125emX}}

\usepackage{verbatim}
\usepackage{balance}
\usepackage{booktabs}
\usepackage{makecell}   
\usepackage{multirow}
\usepackage{xpatch}
\usepackage{xcolor}
\usepackage{colortbl}
\usepackage{amsthm, amssymb}
\usepackage{hyperref}

\newtheorem{assumption}{Assumption}
\newtheorem*{assumption*}{Assumption}

\usepackage{soul}

\newcommand*{\myeqref}[2][Eq.~]{%
  \hyperref[{#2}]{#1(\ref*{#2})}%
}
\def\equationautorefname#1#2\null{%
  Equation#1(#2\null)%
}

\begin{document}
\history{Date of publication xxxx 00, 0000, date of current version xxxx 00, 0000.}
\doi{10.1109/ACCESS.2024.0429000}

\title{Inverse Nonlinearity Compensation of Dielectric Elastomers for Acoustic Actuation}
\author{
    \uppercase{Jin Woo Lee}\authorrefmark{1}, \IEEEmembership{Member, IEEE},
    \uppercase{Gwang Seok An}\authorrefmark{1},
    \uppercase{Jeong-Yun Sun}\authorrefmark{2}, and\\
    \uppercase{Kyogu Lee}\authorrefmark{1,3,4}, \IEEEmembership{Senior Member, IEEE}
}

\address[1]{Department of Intelligence and Information, Seoul National University, Seoul 08826, South Korea}
\address[2]{Research Institute of Advanced Materials (RIAM) and Department of Materials Science and Engineering at Seoul National University, Seoul 08826, South Korea}
\address[3]{Interdisciplinary Program in Artificial Intelligence (IPAI), Seoul National University, Seoul 08826, South Korea}
\address[3]{Artificial Intelligence Institute (AII), Seoul National University, Seoul 08826, South Korea}

\tfootnote{This work was partly supported by Institute of Information \& communications Technology Planning \& Evaluation (IITP) grant funded by the Korea government(MSIT) [NO.2021-0-02068, Artificial Intelligence Innovation Hub (Artificial Intelligence Institute, Seoul National University)] and a National Research Foundation of Korea (NRF) grant funded by the Korean Government (NRF2018M3A7B4089670).}

\markboth
{J. W. Lee \headeretal: Inverse Nonlinearity Compensation of Dielectric Elastomers for Acoustic Actuation}
{J. W. Lee \headeretal: Inverse Nonlinearity Compensation of Dielectric Elastomers for Acoustic Actuation}

\corresp{Corresponding author: Kyogu Lee (e-mail: kglee@snu.ac.kr).}

\begin{abstract}
This paper presents an in-depth examination of the nonlinear deformation induced by dielectric actuation in pre-stressed ideal dielectric elastomers. A nonlinear ordinary differential equation that governs this deformation is formulated based on the hyperelastic model under dielectric stress. By means of numerical integration and neural network approximations, the relationship between voltage and stretch is established. Neural networks are utilized to approximate solutions for voltage-to-stretch and stretch-to- voltage transformations obtained via an explicit Runge-Kutta method. The efficacy of these approximations is illustrated by their use in compensating for nonlinearity through the waveshaping of the input signal. The comparative analysis demonstrates that the approximated solutions are more accurate than baseline methods, resulting in reduced harmonic distortions when dielectric elastomers are used as acoustic actuators. This study highlights the effectiveness of the proposed approach in mitigating nonlinearities and enhancing the performance of dielectric elastomers in acoustic actuation applications.
\end{abstract}


\titlepgskip=-11pt

\maketitle

\section{Introduction}

\PARstart{D}{ielectric} elastomers represent a distinct class of materials renowned for their exceptional response to electrical stimuli, manifesting substantial deformations upon activation.
These elastomers, comprising compliant layers interspersed between electrodes \cite{pelrine2000high}, akin to rubber in their compliance, exhibit remarkable deformability when subjected to an electric field.
The resultant electrostatic force-induced deformation enables the conversion of electrical energy into mechanical motion \cite{pelrine2000high,huang2012giant}.
Composed primarily of compliant elastomeric substances, these materials have garnered significant interest across diverse technological domains, laying the foundation for Dielectric Elastomer Actuators (DEAs).

DEAs, crafted with one or more layers of dielectric elastomeric material sandwiched between compliant electrodes, serve as transducers converting electrical energy into mechanical motion \cite{pelrine2000high}.
Voltage application across the electrodes induces charge accumulation, leading to elastomer layer deformation.
Electrode materials encompass graphite powder, silicone oil, or graphite mixtures, ensuring compliance without constraining elastomer elongation \cite{rogers2013clear}.
These deformations propel the actuator, generating motion or force, albeit necessitating a high threshold voltage for effective transduction \cite{huang2012giant}.
Proper treatments facilitate robust and tunable actuation with improved stretch response rates \cite{kim2019electroactive}.

Initially recognized for remarkable areal expansion in the early 2000s \cite{kornbluh2002dielectric}, DEAs continue to evolve \cite{kim2019electroactive} along with evolutions in the materials as mechanical actuators \cite{kanan2021computational}.
While boasting inherent stretchability, DEAs offer simplicity, cost-effectiveness, lightweight design, and versatile utility \cite{kim2019stretchable}.
Their applications span diverse fields, including loudspeakers \cite{heydt1998design,heydt2000acoustical}, vibration control \cite{herold2011dielectric}, membrane resonators \cite{li2012electromechanical}, noise cancellation, \cite{lu2015tunable,lu2015electronically} and adaptive sound filters \cite{bortot2017tuning}.
The capability to produce substantial deformations and forces while maintaining a lightweight and relatively simple fabrication process underscores their engineering significance.

Employing DEA as an acoustic actuator demands meticulous treatment due to human auditory sensitivity to even minor nonlinear distortions \cite{tan2003effect}, whereas DEA introduces significant nonlinear distortions during voltage-induced elastic deformation \cite{heydt2000acoustical}.
These challenges stem not only from nonlinear elastic deformation \cite{ogden1997non,gent1996new} but also from the nonlinear relationship between voltage and Maxwell stress, even within linear elastic models \cite{pelrine1998electrostriction,heydt2000acoustical}.
Addressing these challenges necessitates controlling nonlinear deformation \cite{liang2024highly}, akin to endeavors focused on enhancing the response characteristics of acoustic actuators.

The dielectric elastomer's deformation exhibits visco-hyperelastic behavior \cite{kanan2021computational}.
Accurate models accounting for large strains and visco-hyperelasticity are imperative for such actuators.
Goulbourne \textit{et al.} \cite{goulbourne2005nonlinear} proposes an analytic solution for dielectric elastomer stretch during actuation, employing Maxwell-Faraday electrostatics and nonlinear elasticity.
While adopting Ogden \cite{ogden1972large} and Mooney-Rivlin \cite{mooney1940theory,rivlin1948large} nonlinear models as approximations to elastic deformation, their solution closely mirrors the DEA model, yet there is potential for enhancement by incorporating more recent elastic models like neo-Hookean \cite{ogden1997non,treloar1943elasticity} or Gent \cite{gent1996new} nonlinear elasticity models.
Modeling viscous behavior necessitates time-varying solutions \cite{xu2012dynamic}.
Recent approaches utilize neural networks to approximate time-dependent DEA dynamics, leveraging recurrent neural networks \cite{zhang2023inverse} or reinforcement learning \cite{li2019deep}.
Irrespective of the dynamics involved, these methodologies aim to model the actuation system, commonly manifested through ordinary or partial differential equations (ODEs/PDEs).

Recent developments exploring the efficacy of neural networks (NNs) to approximate solutions of differential equations have shown promising results.
In the context of function approximation, neural networks have long been recognized as universal approximators \cite{hornik1989multilayer,hornik1991approximation,leshno1993multilayer}.
Additionally, NNs have proven successful in approximating solutions to differential equations \cite{lagaris1998artificial}.
NNs are not limited to solving ODEs alone \cite{lagaris1998artificial,chen2018neural}; they are applied extensively to tackle PDEs, including the Korteweg-de Vries (KdV) equation \cite{li2021fourier,gupta2021multiwavelet}, or Navier-Stokes equations \cite{lee2019data,raissi2019physics}.

This paper mitigates the problem of compensating for the nonlinear deformation caused by dielectric actuation using NNs.
In particular, the focus is on reducing the computational overhead required by the numerical solver by utilizing NNs, while also identifying improved compensation performance through training methods that utilize automatic differentiation and gradient backpropagation.
We introduce the literature that is relevant to what we will cover in this paper (\autoref{sec: related}).
Then we formalize the problem of interest (\autoref{sec: problem}).
We deduce a nonlinear ODE using the hyperelastic energy inherent in an ideal DE-based system (\autoref{sec: simulating}).
Subsequently, we establish the relationship between voltage and stretch using both numerical integration methods and neural network approximations (\autoref{sec: modeling}).
Leveraging the approximated solution, we apply it to compensate for the nonlinear dielectric elastomer actuation.
Our evaluation in \autoref{sec: evaluation} demonstrates the effectiveness of these methods.

\begin{figure*}
    \includegraphics[width=\textwidth]{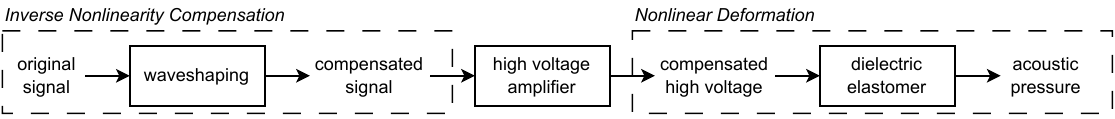}
    \caption{\label{fig: overview}Overview of the system.}
\end{figure*}

\section{Related Works}\label{sec: related}

\subsection{Modeling Hyperelastic Deformation}
The complex behavior of elastomers necessitates the use of hyperelastic theories to derive the elastic energy of DE-based systems.
Multiple models elucidate hyperelasticity, such as the neo-Hookean \cite{treloar1943elasticity,ogden1997non}, Gent \cite{gent1996new}, Mooney-Rivlin \cite{mooney1940theory,rivlin1948large}, Ogden \cite{ogden1972large}, and Arruda-Boyce \cite{arruda1993three}.
Recent studies have been dedicated to understanding the behaviors of DEAs.
For instance, Jia \textit{et al.} \cite{jia2019deformation,li2012electromechanical} conducts an analytical study on a flat DE membrane's in-plane deformation, while \cite{zhu2010resonant,garnell2019dynamics} explores the nonlinear dynamics of a balloon-shaped DE, revealing (sub-)harmonic responses under specific actuation conditions.
This paper is distinguished by the fact that unlike the former \cite{jia2019deformation,li2012electromechanical}, it analyzes an idealized DEA that does not consider viscous effects, and unlike the latter \cite{zhu2010resonant,garnell2019dynamics}, it analyzes a planar-shaped membrane.
Additionally, this study aligns with literature analyzing the hyperelasticity using neo-Hookean model \cite{xu2012dynamic,zhang2016dynamic}, which is adopted within strains without reaching limiting stretch \cite{treloar1943elasticity,gent1996new}.

\subsection{Inverse Nonlinearity Compensation}\label{ssec: related inc}
Leveraging the insights from modeling hyperelastic deformation, our objective is to develop compensation techniques that reverse the nonlinear forward model.
In the context of control theory, similar problems have been tackled through inverse dynamics compensation (IDC) \cite{miyamoto1988feedback}.
IDC involves calculating and applying controls to counteract disturbances or achieve precise movements in dynamic systems, frequently employed in robotics or biomechanics.
For instance, \cite{zou2018high,zou2019feedforward} successfully applied IDC to manage viscous hysteresis in DEAs using experimentally measured data.
Analogous to IDC, this paper focuses on processing input signals to nonlinear DEAs to attain linear actuation.
We frame this as an Inverse Nonlinearity Compensation (INC) problem aimed at controlling hyperelastic deformation induced by the Maxwell stress \cite{jastrze2009compensation,skricka2002improvements}.
Considering the function's nonlinear and time-invariant nature, we categorize this inverse nonlinearity compensation process as a form of \textit{waveshaping}.

\subsection{Neural Waveshaping Functions}\label{ssec: related neural}
It is pertinent to acknowledge the conventional notion of waveshaping as a synthesis method \cite{le1979digital}, designed to generate timbres from sinusoidal inputs using a memoryless, shift-invariant nonlinear shaping function \cite{arfib1978digital,dodge1985computer}.
This function introduces harmonic components to the input signal \cite{roads1979tutorial}.
Recently, \cite{hayes2021neural} proposes a neural waveshaping unit (NEWT) that enhances a harmonic-plus-noise synthesizer \cite{serra1990spectral} with neural networks, akin to Differentiable Digital Signal Processing (DDSP) \cite{engel2020ddsp}.
NEWT employs Multi-Layer Perceptrons (MLPs) to learn continuous representations of waveshaping functions and perform comparably to DDSP in synthesis tasks, doing so more efficiently.
Our method shares the same methodology as NEWT by implicitly learning the waveshaping function using a neural network.
However, it's crucial to highlight that we're not engaged in a synthesis task.
Specifically, the purpose of the waveshaping function in our context is diametrically opposite to conventional waveshaping; it aims to \textit{eliminate} the harmonics rather than induce them.
This distinction will be rigorously addressed in the subsequent section.

\section{Problem Statement}\label{sec: problem}
\begin{figure}
    \begin{minipage}[b]{0.39\linewidth}
        \centering
        \centerline{\includegraphics[width=\linewidth]{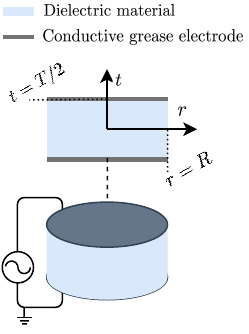}}
        \vspace{-2mm}
        \centerline{\footnotesize (a) Reference state}
    \end{minipage}
    \begin{minipage}[b]{0.59\linewidth}
        \centering
        \centerline{\includegraphics[width=\linewidth]{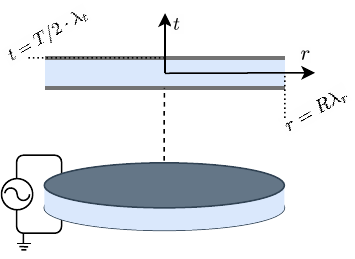}}
        \vspace{-2mm}
        \centerline{\footnotesize (b) Deformed state}
    \end{minipage}
    \caption{Schematic diagram of dielectric system cross-section.
    (a) Reference state where no external force is applied.
    (b) Deformed state with equi-biaxial tension.
    }
    \label{fig: dea}
\end{figure}

Denote the stretch in DEA as $\lambda$.
This is determined by both pre-stretch and electromagnetic deformations influenced by the amplified high voltage signal. Denote the amplification factor and the signal by $\alpha$ and $x$, respectively, so that the high voltage signal is expressed as ${V}=\alpha\cdot{x}$.
In this paper, equi-biaxial deformation, as depicted in \autoref{fig: dea}, is considered and the pre-stretch is applied in the same direction.
This study focuses on two main aspects: modeling the system and controlling it.
First, we construct a model of the voltage-stretch system using an ODE.
By solving this ODE, we obtain a solution that relates voltage to stretch.
\begin{align}
    f: &\ \mathbb{R} \to \mathbb{R}, \\
       &\ {V} \mapsto \lambda
\end{align}
Subsequently, we aim to find a signal-mapping function:
\begin{align}
     g: &\ \mathbb{R} \to \mathbb{R}, \\
        &\ {x} \mapsto \hat{{x}},
\end{align}
such that the resulting DEA stretch $\hat{\lambda}=f(\alpha\cdot\hat{{x}})$ maintains a linear correlation with the original signal ${x}$.
Function $g$ serves as an inverse nonlinearity compensator of $f$ if for some $a_0,a_1\in\mathbb{R}$ with $a_1\neq0$,
\begin{equation}\label{eqn: inc}
    f(\alpha\cdot g({x}))=a_0 + a_1{x}.
\end{equation}
Upon determining the fixed $f$ and its corresponding compensator $g$, this paper proposes approximated solutions $f_\theta$ and $g_\theta$, parameterized by $\theta$, optimized to:
\begin{align}\label{eqn: objective}
    \text{minimize}&\quad \mathbb{E}_{V}\Big[\vert f_\theta({V}) - f({V})\vert \Big] + \mathbb{E}_{x}\Big[\vert g_\theta({x}) - g({x})\vert \Big], \\
    \text{subject to}&\quad {x}>0\quad\text{and}\quad {V} = \alpha{x}>0.
\end{align}
The subscript $\theta$ indicates that $f$ and $g$ each have trainable parameters, which have independent parameter updates.
The subsequent sections delve into the problem statement.
However, for a more structured approach, we initially derive the ODE model, propose an approximation method for its solutions $f$ and $f^{-1}$, and then demonstrate that the approximation alone satisfactorily addresses problem \autoref{eqn: objective}.

\section{Simulating Dielectric Elastomer Actuator}\label{sec: simulating}
The schematic overview of the proposed system is illustrated in \autoref{fig: overview}.
Our focus initiates with the modeling of the voltage-stretch system, specifically delving into the electromechanical actuation of DEAs.
We begin by defining the Helmholtz free energy of the elastomer as $A$.
The potential energy $U$ of the dielectric elastomer is structured as per the formulations in \cite{wang2016nonlinear, alibakhshi2019analytical}:
\begin{equation}\label{eqn: potential energy}
    U = A - W_V - W_F
\end{equation}
Here, $A$ denotes the Helmholtz free energy, while $W_V$ represents the work done by the voltage, and $W_F$ depicts the work done by the tensile force.
Applying the difference operator to \autoref{eqn: potential energy} yields:
\begin{equation}\label{eqn: free energy}
    \delta A = {V} \delta Q + F_r \delta\lambda_r
\end{equation}
Defining the nominal density of the free energy as $W = A / (\pi R^2T)$, \autoref{eqn: free energy} can be reformulated as:
\begin{equation}\label{eqn: energy conservation}
    \pi R^2T \delta W = {V} \delta Q + F_r \delta\lambda_r.
\end{equation}
To further progress and derive an ODE for ${V}$ and $\lambda$, we express $\delta W$ and $\delta Q$ of \autoref{eqn: energy conservation} in terms of $\delta\lambda$.
Before proceeding, we introduce specific assumptions in the actuation process to streamline the derivation of the equation of motion.
\begin{assumption}[No polarization]\label{assm: no polarization}
There exists no polarization between the electrodes. Hence, the electric displacement field is defined as $D=\varepsilon E$, where $\varepsilon$ and $E$ represent the dielectric constant and the electric field, respectively.
\end{assumption}
\begin{assumption}[Incompressibility]\label{assm: incompressibility}
The dielectric elastomer is considered incompressible, signified by $J = \lambda_r \lambda_\theta \lambda_t = 1$. Here, subscripts $r$, $\theta$, and $t$ refer to directions in cylindrical coordinates for radius, angle, and thickness.
\end{assumption}
\begin{assumption}[No buckling]\label{assm: no buckling}
The stretch along the angular direction $\lambda_\theta$ remains constant at a value of $1$ for all $\theta$, ensuring the absence of buckling in the dielectric elastomers.
\end{assumption}
Additionally, a free boundary condition is assumed, without support or clamping, and the electrode's mass is considered negligibly small.

\subsection{Electric Displacement}
To reformulate $\delta Q$ in \autoref{eqn: energy conservation}, we initially establish the relationship between the electric displacement and the electric field as per Assumption \ref{assm: no polarization}.
Given the definitions of the electric field $E={V}/t$ and the electric displacement field $D=Q/(\pi r^2)$, we derive:
\begin{equation}
    \frac{Q}{\pi r^2} = \varepsilon\frac{{V}}{t}
\end{equation}
Here, $r$ and $t$ represent the radius and thickness of the dielectric material in a deformed state, respectively.
$\varepsilon=\varepsilon_0\varepsilon_r$ denotes the permittivity, where $\varepsilon_0$ signifies the vacuum permittivity and $\varepsilon_r$ stands for the relative permittivity (or the dielectric constant).
By rewriting $r=R\lambda_r$ and $t=T\lambda_t$, we arrive at the following expression:
\begin{equation}\label{eqn: Q1}
    Q = \varepsilon{V}\frac{\pi r^2}{t} = \varepsilon{V}\frac{\pi R^2}{T}\frac{\lambda_r^2}{\lambda_t}
\end{equation}
Given Assumptions \ref{assm: incompressibility} and \ref{assm: no buckling}, where $\lambda_r\lambda_t=1$, \autoref{eqn: Q1} simplifies to:
\begin{equation}\label{eqn: Q2}
    Q = {V}\frac{\pi R^2}{T}\frac{\varepsilon}{\lambda_t^2}.
\end{equation}
Utilizing the difference operator $\delta\cdot$, we can express $\delta Q$ in terms of $\delta{V}$ and $\delta\lambda_t$:
\begin{equation}\label{eqn: delta Q}
    \delta Q = \delta{V}\frac{\pi R^2}{T}\frac{\varepsilon}{\lambda_t^2} - 2{V}\frac{\pi R^2}{T}\frac{\varepsilon}{\lambda_t^3}\cdot\delta\lambda_t
\end{equation}
Substituting the $\delta Q$ using \autoref{eqn: delta Q} further simplifies the \autoref{eqn: energy conservation}.

\subsection{Hyperelastic Deformation}
To proceed, it remains to rephrase $\delta W$ in \autoref{eqn: energy conservation} in terms of $\delta\lambda_t$, integrating elastic deformation models into the analysis.
Commencing with the free energy density originating from the neo-Hookean model \cite{treloar1943elasticity,ogden1997non}, it is represented as:
\begin{equation}\label{eqn: neo-hookean free energy}
    W = \frac{\mu}{2}(I-3)
\end{equation}
Here, $I = \lambda_r^2 +\lambda_\theta^2 +\lambda_t^2$ characterizes the trace of the right Cauchy-Green deformation tensor.
With Assumptions \ref{assm: incompressibility} and \ref{assm: no buckling}, the expression $I$ can be redefined as:
\begin{equation}
    I = \lambda_t^2 + 1 + \lambda_t^{-2}.
\end{equation}
Consequently, the neo-Hookean free energy density in \autoref{eqn: neo-hookean free energy} transforms into:
\begin{equation}
    W = \frac{\mu}{2}(\lambda_t^2 + \lambda_t^{-2} - 2)
\end{equation}
Employing the difference operator allows us to express $\delta W$ of the neo-Hookean model using $\delta\lambda_t$:
\begin{equation}\label{eqn: delta W}
    \delta W = \mu(\lambda_t -\lambda_t^{-3})\delta \lambda_t
\end{equation}
Finally, the reconfiguration of \autoref{eqn: energy conservation} using \autoref{eqn: delta Q} and \autoref{eqn: delta W} results in the following ODE:
\begin{equation}\label{eqn: ode}
    \frac{\delta{V}}{\delta\lambda_t} = 2\frac{{V}}{\lambda_t} + \frac{\mu T^2}{\varepsilon{V}}(\lambda_t^{-1} - \lambda_t^{3}) - \frac{F_rT}{\pi R^2}\cdot\frac{1}{\varepsilon{V}}
\end{equation}
This equation encapsulates the relationship between the driving voltage and the stretch within the dielectric elastomer model, providing a foundation for further analysis.

\section{Modeling the Voltage-Stretch Relation}\label{sec: modeling}
To establish the relationship between voltage and stretch, it is crucial to solve the ODE in \autoref{eqn: ode}.
Solving \autoref{eqn: ode} involves addressing this Initial Value Problem (IVP) through the use of numerical integration techniques.
As the initial value, we set the reference state where $\lambda_t=1$ with $V=0$.

\begin{figure}
    \centering
    \begin{minipage}[b]{0.32\linewidth}
        \centering
        \centerline{\includegraphics[width=\linewidth]{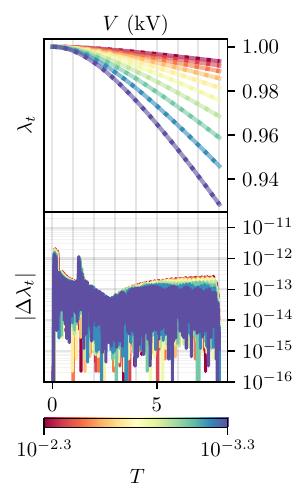}}
        \centerline{\footnotesize (a) Thickness $T$}
    \end{minipage}
    \begin{minipage}[b]{0.32\linewidth}
        \centering
        \centerline{\includegraphics[width=\linewidth]{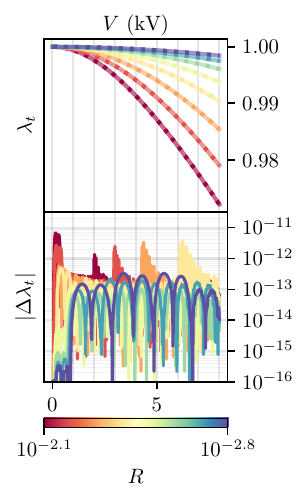}}
        \centerline{\footnotesize (b) Radius $R$}
    \end{minipage}
    \begin{minipage}[b]{0.32\linewidth}
        \centering
        \centerline{\includegraphics[width=\linewidth]{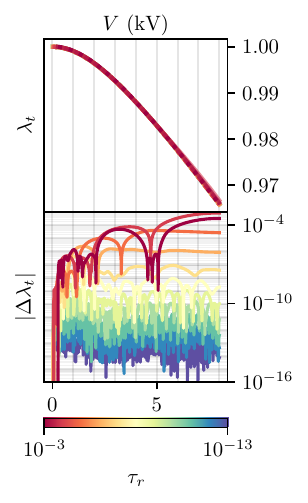}}
        \centerline{\footnotesize (c) Relative tolerance $\tau_r$}
    \end{minipage}
    \caption{\label{fig: span}{Voltage-stretch curve obtained using RK45.
    The voltage-to-stretch $f$ curves are plotted in solid lines and the stretch-to-voltage $f^\dagger$ curves are plotted in dotted lines.
    Colors distinguish variations in $T$, $R$, and $\tau_r$.
    }}
\end{figure}


\subsection{Approximation using Numerical Integration Method}\label{ssec: approximation using numerical integration}
First, the solution is obtained by solving IVP using a numerical integration method.
An explicit Runge-Kutta method of order 5(4) \cite{dormand1980family}, denoted as RK45, is employed as the numerical integrator.

To obtain the voltage-to-stretch solution \(f\), we initialize the process with \(\lambda_t[0]=1\) and \(V[0]=0\), and then iteratively determine \(\lambda_t[n]\) using an integrator while varying \(V[n]\) for \(n \in \mathbb{Z}\).
The stretch-to-voltage solution $f^{-1}$ is obtained through a similar procedure.
Utilizing the RK45 integrator necessitates defining tolerances, as with many other numerical integration techniques.
These tolerances, comprising an absolute tolerance (\(\tau_a\)) and a relative tolerance (\(\tau_r\)), serve to delineate the upper bound of local error estimates.
For instance, the voltage-to-stretch solver continues iterating until the solution converges, ensuring its error remains below \(\tau_a + \tau_r \cdot |\hat{\lambda}_t|\).
Conversely, the stretch-to-voltage process adheres to a similar criterion with \(\tau_a + \tau_r \cdot |\hat{V}|\) for convergence.

\autoref{fig: span} (a) and (b) illustrate the solutions of \autoref{eqn: ode} corresponding to various $T$ and $R$.
Given that the stretch-to-voltage functions analyzed here approximate the inverse function of $f$, we use the notation $f^\dagger$ instead of $f^{-1}$.
In each experiment, while varying the $T$ and $R$ values within their respective ranges, the other mechanical properties maintain the values detailed in \autoref{tab: params}.
For reference, specific values such as $T=10^{-3.3}\approx5\times10^{-4}$ and $T=10^{-2.3}\approx5\times10^{-2}$ cover typical initial thicknesses of VHB\textsuperscript{\texttrademark} series, while $R=10^{-2.8}\approx1.5\times10^{-3}$ and $R=10^{-2.1}\approx8\times10^{-3}$ cover typical initial radii of the DEAs.
The subplots (a) and (b) are simulated with $\tau_a=10^{-16}$ and $\tau_r=10^{-13}$.
The numerical integrator consistently produces precise solutions for various parameters, with $f$ and $f^\dagger$ exhibiting errors less than $10^{-11}$.

\autoref{fig: span} (c) shows the solutions corresponding to various tolerance $\tau_r$.
The absolute tolerance values are set $\tau_a=10^{-3}\cdot\tau_r$.
Since the tolerance determines the error of the solution, the figure shows that the difference between $f$ and $f^\dagger$ increases as the tolerance increases.
However, enhancing efficiency remains a potential area for improvement, as the current approach necessitates a recursive update of solutions through numerical integration from the initial value to each target point.
This leads to an approximated solution in an implicit form, estimating the output without the need for iteration or integration at each step.

\autoref{tab: params} presents the notation and associated values for each parameter, corresponding to \textsc{3M}\textsuperscript{\texttrademark} \textsc{VHB}\textsuperscript{\texttrademark} Tape 4910.
The specific values align with Young's modulus and permittivity, with detailed references available in \cite{bozlar2012dielectric} and \cite{vu2012impact}.

\begin{table}[t]
    \centering
    \caption{\label{tab: params} Mechanical properties}
    \begin{tabular}{llll}\toprule
        Notation & Value & Unit & Descriptions \\\midrule
        $T$ & $10^{-3}$ & m & Initial thickness \\
        $R$ & $10^{-2}$ & m & Initial radius \\
        $Y$ & $220$ & kPa & Young's modulus \\
        $\varepsilon_0$ & $8.85\times10^{-12}$ & $\text{C}^2/\text{m}^{2}\text{N}$ & Vacuum permittivity \\
        $\varepsilon_r$ & $3.5$ & - & Relative permittivity
    \\\bottomrule\end{tabular}
\end{table}

\subsection{Approximation using Neural Network}

\begin{figure}
    \centering
    \includegraphics[width=0.8\linewidth]{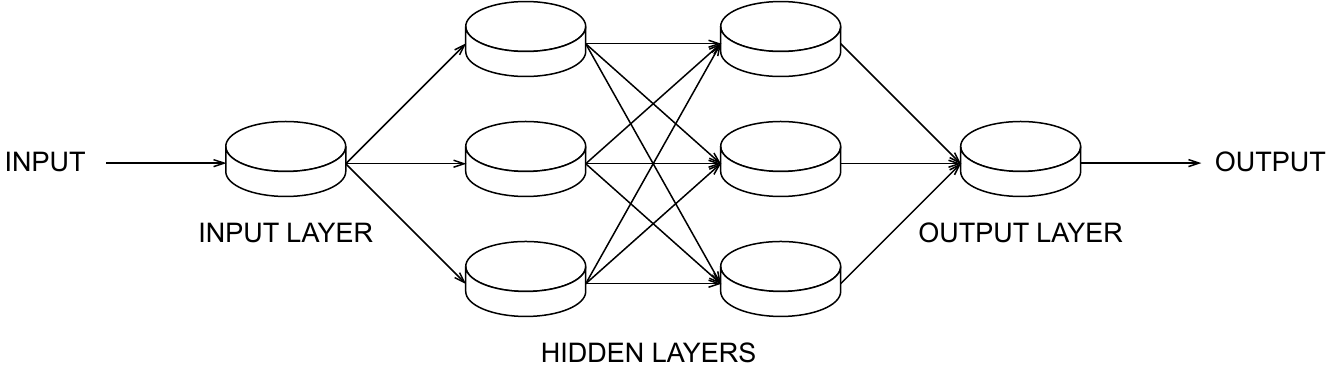}
    \caption{\label{fig: mlp}{
        Example illustration of a multi-layer perceptron. Each bundle of arrows represents a linear transformation, and the nodes in each layer represent pointwise nonlinearities.
    }}
    \vspace{-5mm}
\end{figure}

The methodology explored in \autoref{ssec: approximation using numerical integration} delineates a nonlinear yet time-invariant solution.
Although several methods exist to approximate such functions --- such as Taylor series expansion --- we propose an implicit approximation through neural networks.
In our investigation, we opt for an MLP as a specific choice to approximate a memoryless system.
In consonance with the conceptual ethos presented in NEWT \cite{hayes2021neural}, we engineer the MLP with three layers and employ sinusoidal activation functions between these layers.
The network architecture follows that of the conventional MLPs, similar to the system depicted in \autoref{fig: mlp}.
The MLP is trained by simulating the voltage-stretch pair using the RK45 integrator.
The voltage values within the training set are uniformly sampled from $V_{\text{min}}=0$ kV to $V_{\text{max}}=8$ kV.
Each training iteration involves sampling voltage-stretch pairs, aggregating to a total batch size of $N=1024$.

\begin{figure}
    \centering
    \begin{minipage}[b]{0.30\linewidth}
        \centering
        \centerline{\includegraphics[width=\linewidth]{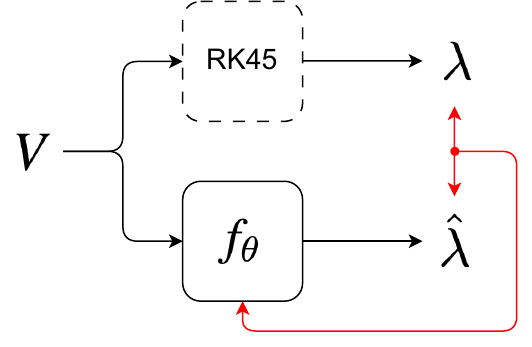}}
        \centerline{\footnotesize (a) Standard training of $f_\theta$}
    \end{minipage}
    \begin{minipage}[b]{0.50\linewidth}
        \centering
        \centerline{\includegraphics[width=\linewidth]{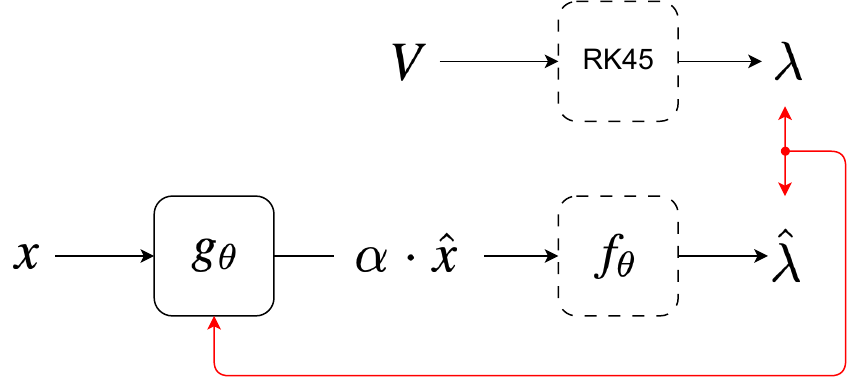}}
        \centerline{\footnotesize (b) E2E training of $g_\theta$}
    \end{minipage}
    \caption{\label{fig: training}{
        Illustration of training processes.
        Modules where there are no trainable parameters or where parameters are not updated are marked with dashed lines.
        Red solid lines indicate the gradient back-propagation through the measurement of losses.
        Please note that $g_\theta$ can also be trained through a standard training process by measuring the loss between $V$ and $\alpha\cdot\hat{x}$.
    }}
    \vspace{-5mm}
\end{figure}

The training procedures for $f_\theta$ and $g_\theta$ exhibit a slight variance: while $f_\theta$ is directly updated by gauging the loss using its output, $g_\theta$ is updated by measuring the loss after its amalgamation with $f_\theta$ as $f_\theta\circ g_\theta$.
\autoref{fig: training} shows the schematics of the training processes.
This leverages the differentiable property of $f_\theta$, thereby employing the pre-trained and fixed $f_\theta$ to facilitate an end-to-end (E2E) training for $g_\theta$.
For training $f_\theta$, labeled pairs $\mathbf{V},\mathbf{\Lambda}$ are imperative.
Therefore, we uniformly sample $\mathbf{V}$ within the range from $V_\text{min}$ to $V_\text{max}$ and simulate $\mathbf{\Lambda}$ using RK45 in each iteration.
Conversely, as the objective for $g_\theta$ primarily involves learning to invert $f_\theta$, such simulation is deemed unnecessary for training $g_\theta$.
Consequently, the training of $g_\theta$ omits the RK45 scheme, leading to nearly a twofold increase in training speed.

\section{Evaluation}\label{sec: evaluation}
This section aims to assess the validity of the approximation as an inverse nonlinearity compensator $g$, also denoted as the waveshaping function in \autoref{fig: overview}.
Converting $f^{-1}$ into the waveshaping function is straightforward.
Given our knowledge of $f$ and its inverse $f^{-1}$, the inverse nonlinearity compensator of $f$ can be represented by
\begin{equation}\label{eqn: g}
    g(x) = \frac{1}{\alpha}f^{-1}(a_0+a_1x)
\end{equation}
as defined in \autoref{eqn: inc}.
Here, $f$ simulates the stretch for a given voltage, while $f^{-1}$ determines the driving voltage required to achieve the desired stretch.
Thus, rescaling the signal $x \in [-1,1]$ to the stretch $\lambda_t \in [\lambda_{\text{floor}}, \lambda_{\text{ceil}}] \subset (0,1]$ suffices obtaining the inverse nonlinearity compensator $g$, as $f^{-1}$ indicates the required voltage for the specified stretch.

The affine transformation by $a_0$ and $a_1$ rescales $x \mapsto \lambda_t$, where $a_0$ and $a_1$ are determined by the driving voltage range.
Fixing the AC and DC components of the driving voltage ($V_{\text{pp}}$ and $V_{\text{dc}}$, respectively), where $V_{\text{min}} \leq V_{\text{floor}} = V_{\text{dc}} - V_{\text{pp}}/2$ and $V_{\text{dc}} + V_{\text{pp}}/2 = V_{\text{ceil}} \leq V_{\text{max}}$, sets the range of stretch as $\lambda_{\text{floor}} = f(V_{\text{floor}})$ and $\lambda_{\text{ceil}} = f(V_{\text{ceil}})$, along with determining $a_0$ and $a_1$.
To assess how accurate the approximation $f^\dagger$ is in comparison to $f^{-1}$, we evaluate their proximity in the context of the INC, by replacing $f^{-1}$ in \autoref{eqn: g} with $f^\dagger$.
For the particular choice of the coefficients, we refer the readers to Appendix \ref{ssec: inc function}.

\subsection{Baselines}\label{ssec: baselines}
Prior research aiming to mitigate acoustical distortion in DEAs has often relied on the square root function \cite{heydt2000acoustical}.
When considering the linear elastic model, employing the square root on the voltage signal is intuitive.
However, in our case of nonlinear elastic deformation, more suitable adaptations are necessary.
Hence, we propose to benchmark our approach against two baselines: a trainable power function and a piecewise polynomial interpolation.

\subsubsection{Power Function Fitting}\label{sssec: power}
Our first baseline involves a power function, where the power exponent becomes a trainable parameter.
This can be viewed as an extension of the square-root compensation advocated by \cite{heydt2000acoustical}.
The power function, characterized by fitted parameters $\theta_{j}^{(1)}$, is articulated as follows:
\begin{equation}\label{eqn: pff}
    f^{(1)} = \theta_{1}^{(1)} \left(|\theta_{2}^{(1)}(1-\lambda_t)+\theta_{3}^{(1)}|\right)^{\theta_{0}^{(1)}}  + \theta_{4}^{(1)}
\end{equation}
Affine transformations precede and succeed in the power operation to fine-tune the scale between stretch and voltage.
To ensure numerical stability, an absolute operator ($|\cdot|$) is applied before the power operation.
We denote the \autoref{eqn: pff} as Power Function Fitting (PFF).
For a more comprehensive understanding of the other families of trainable functions as baselines, we direct interested readers to Appendix \ref{sec: appendix baseline details}.

\subsubsection{Piecewise Quartic Interpolation}\label{sssec: polynomial}
Long-standing efforts have been made to interpolate Runge-Kutta solutions \cite{horn1983fourth,shampine1985interpolation}.
From this line of work, we compare our method with a Piecewise Quartic Interpolation (PQI) \cite{shampine1986some}.
The interpolation is performed in a piecewise manner, wherein each segment of the solution is approximated using a quartic polynomial.
While it is possible to tune the accuracy of PQI by adjusting the tolerance in the RK45, we fix $\tau_r = 10^{-6}$ and $\tau_a = 10^{-9}$ for evaluation.
The selected values are thoughtfully curated to ensure an equitable comparison, taking into account computational expenses (refer to \ref{ssec: computational efficiency} for further details).

\begin{figure}[t]
    \centering
    \includegraphics[width=0.8\linewidth]{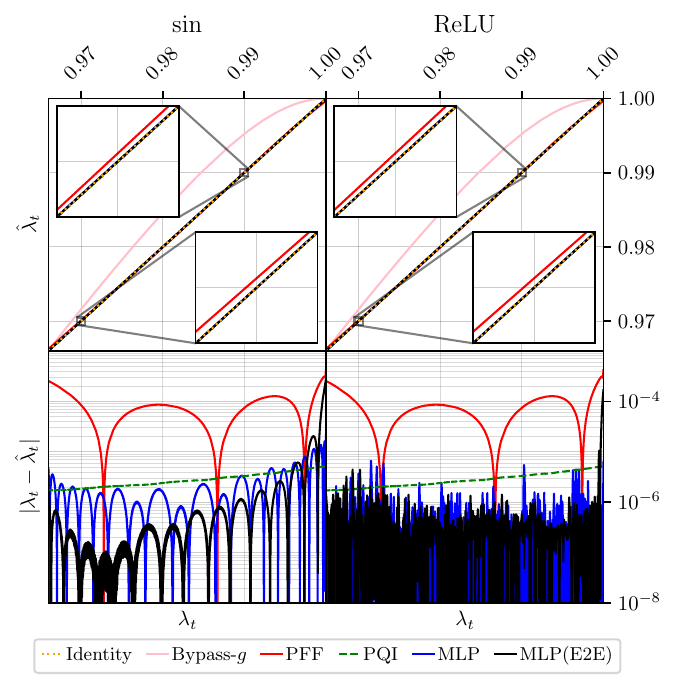}
    \caption{\label{fig: id}{Plot of $\hat{\lambda}_t=f\circ f^{\dagger}$ and $|\lambda_t-\hat{\lambda}_t|$ against $\lambda_t$.
    }}
\end{figure}

\subsection{Evaluation metric}\label{ssec: evaluation metric}
We assess the outcomes using three key metrics: log of the $\ell_1$-norm of stretch difference ($\log\ell_1$), $\ell_1$-norm of the log magnitude spectrogram difference ($\ell_1^\text{STFT}$), and Signal-to-Distortion Ratio (SDR).
The log of the $\ell_1$-norm of stretch difference is calculated as
\begin{equation}
    \log\ell_1(\hat{\lambda}_t,\lambda_t) := \log_{10}\|\hat{\lambda}_t-\lambda_t\|_1
\end{equation}
where $\hat{\lambda}t=f(\hat{x})$ represents the estimated stretch of the compensated signal $\hat{x}$.
To calculate $\log\ell_1$, we linearly probe $x$ within the range $x\in[V_{\text{min}},V_{\text{max}}]$.
However, the remaining metrics are applied to evaluate sinusoidal signals specifically.
For instance, the sinusoid $x[n]=V_{\text{dc}}+(V_\text{pp}/2)\sin(\omega n)$ with $V_{\text{dc}}=6$ kV and $V_{\text{pp}}=3$ kV varies its frequency within the range between 0 to $12$ kHz.
The sampling rate $f_s$ is set to 48 kHz.
Upon obtaining the estimated $\hat{\lambda}_t[n]=f(\hat{x}[n])$ for the sinusoidal $\hat{x}$, we compare the log magnitude difference of the spectrograms as
\begin{equation}
    \ell_1^\text{STFT}(\hat{\lambda}_t,\lambda_t) := \left\|10\log_{10}\left|\frac{\text{STFT}(\hat{\lambda}_t)}{\text{STFT}(\lambda_t)}\right|^2 \right\|_1
\end{equation}
and the signal-to-distortion ratio as
\begin{equation}
    \text{SDR}(\hat{\lambda}_t,\lambda_t) := 10\log_{10}\frac{\|\lambda_t\|_2^2}{\|\hat{\lambda}_t-\lambda_t\|_2^2}
\end{equation}
which represents the ratio of $\ell_2$-norms between the reference signal $\lambda_t$ and the distortion $\hat{\lambda}_t-\lambda_t$ in dB scale.
Among the three indicators, SDR is the only metric where a larger value indicates superior performance, while for the rest, a smaller value signifies better performance.

\begin{figure}[t]
    \centering
    \begin{minipage}[b]{0.23\linewidth}
        \centering
        \centerline{\includegraphics[width=\linewidth]{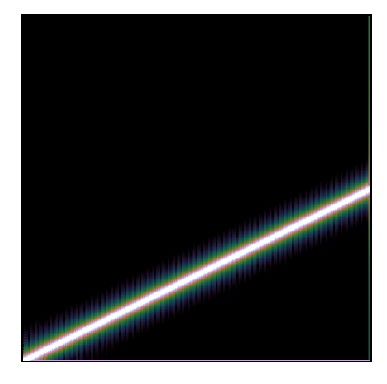}}
        \vspace{-2mm}
        \centerline{\footnotesize (a) Original}
    \end{minipage}
    \begin{minipage}[b]{0.23\linewidth}
        \centering
        \centerline{\includegraphics[width=\linewidth]{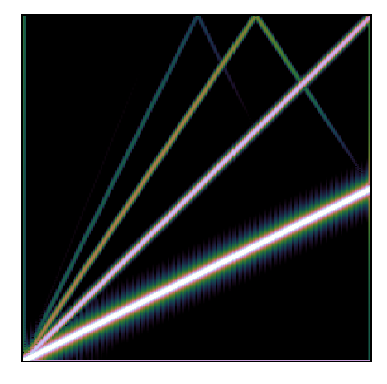}}
        \vspace{-2mm}
        \centerline{\footnotesize (b) Bypass-$g$}
    \end{minipage}
    \begin{minipage}[b]{0.23\linewidth}
        \centering
        \centerline{\includegraphics[width=\linewidth]{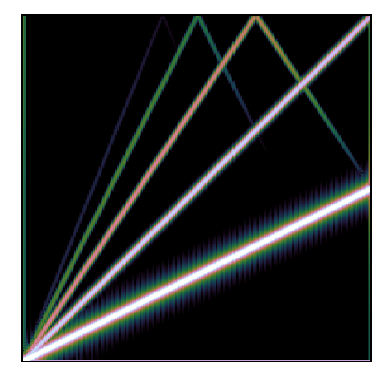}}
        \vspace{-2mm}
        \centerline{\footnotesize (c) HFF}
    \end{minipage}
    \begin{minipage}[b]{0.23\linewidth}
        \centering
        \centerline{\includegraphics[width=\linewidth]{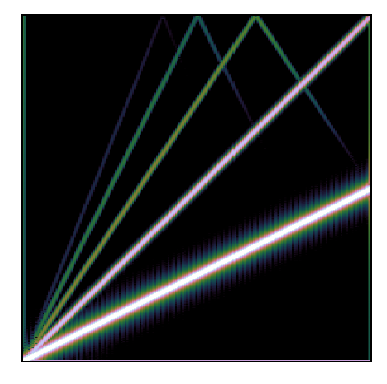}}
        \vspace{-2mm}
        \centerline{\footnotesize (d) LFF}
    \end{minipage}
    \begin{minipage}[b]{0.23\linewidth}
        \centering
        \centerline{\includegraphics[width=\linewidth]{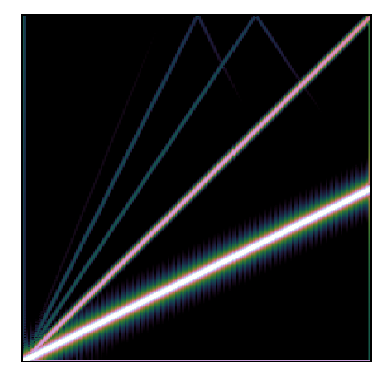}}
        \vspace{-2mm}
        \centerline{\footnotesize (e) PFF}
    \end{minipage}
    \begin{minipage}[b]{0.23\linewidth}
        \centering
        \centerline{\includegraphics[width=\linewidth]{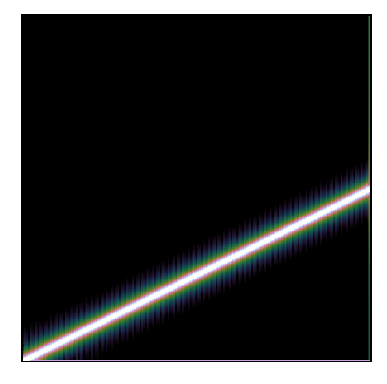}}
        \vspace{-2mm}
        \centerline{\footnotesize (f) PQI}
    \end{minipage}
    \begin{minipage}[b]{0.23\linewidth}
        \centering
        \centerline{\includegraphics[width=\linewidth]{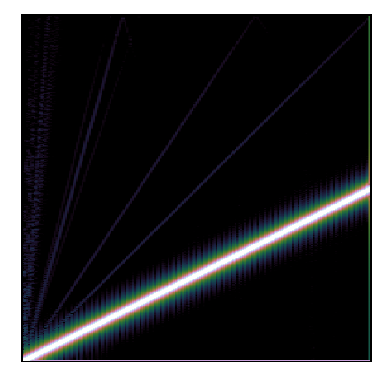}}
        \vspace{-2mm}
        \centerline{\footnotesize (g) MLP(ReLU)}
    \end{minipage}
    \begin{minipage}[b]{0.23\linewidth}
        \centering
        \centerline{\includegraphics[width=\linewidth]{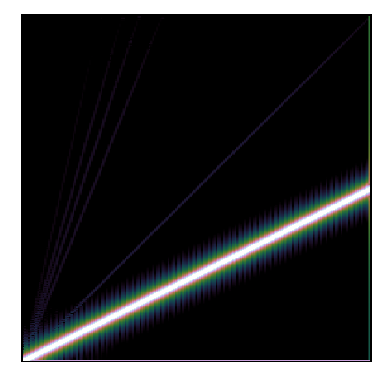}}
        \vspace{-2mm}
        \centerline{\footnotesize (h) MLP(sin)}
    \end{minipage}
    \caption{\label{fig: spec}{
        Spectrograms of the experimented sinusoidal signals.
        Patch (a) shows the spectrogram of the original input as the pure sinusoidal sweep signal.
        Patches (b)-(h) show the spectrograms of the simulated output signals, where (b) shows the output without any compensation and patches (c)-(h) show the outputs with different inverse nonlinearity compensators.
    }}
    \vspace{-6mm}
\end{figure}

\subsection{Evaluation results}\label{ssec: results}
\autoref{fig: id} illustrates the waveshaping function $\alpha g(x)=f^\dagger(a_0+a_1x)$ combined with the deformation function $f$.
The case where no compensation is applied is labeled as Bypass-$g$.
For all $f^\dagger$ models, the fixed forward deformation function $f$ adheres to PQI with $\tau_r=10^{-13}$.
A detailed discussion concerning this choice is available in Appendix \ref{ssec: forward deformation model}.

Each column showcases diverse results for nonlinear activations in MLPs, utilizing sin and ReLU activations for the left and right columns, respectively.
The top row illustrates $f\circ f^{\dagger}(\lambda_t)$ against $\lambda_t$, portraying the proximity to the identity mapping (indicated by {Identity) and indicating the closeness of $f^\dagger\approx f^{-1}$.
Bypass-$g$, where no compensation occurs ($f^\dagger=\textbf{1}$), is plotted for a reference.
Although PFF exhibits compensation performance yielding $f\circ f^\dagger$ close to the identity, the improvement is marginal compared to that of PQI or MLP.
The subsequent bottom row in \autoref{fig: id} shows the errors $|\lambda_t-f\circ f^{\dagger}(\lambda_t)|$, represented on a logarithmic scale.

While PQI consistently demonstrates errors in $|\lambda_t - f \circ f^{\dagger}(\lambda_t)|$ against $\lambda_t$, its average error surpasses that of MLPs.
However, PQI exhibits fewer harmonics, particularly when compared to MLP(ReLU).
This divergence becomes more pronounced when examining the scores in \autoref{tab: results}: MLP(ReLU) outperforms PQI in $\log\ell_1$ with a margin of 0.34, but displays inferior performance in $\ell_1^{\text{STFT}}$ and SDR.
We attribute this phenomenon carefully to the consistency in the error.

An ablation study on the training method is conducted, where the MLP is directly trained from data simulated using RK45 (labeled MLP).
Conversely, the proposed MLP(E2E) demonstrates superior performance on average.
Nevertheless, MLP(E2E) with sin activation exhibits reduced accuracy in small strain regions ($\lambda_t>0.99$).
In contrast, MLP(E2E) with ReLU activation shows the best $\log\ell_1$ score (as detailed in \autoref{tab: results}).
However, ReLU introduces non-smoothness, evident in the jagged error plot, warranting a deeper examination through objective evaluation scores.

\begin{table}[t]
    \centering
    \caption{\label{tab: results}{Evaluation results}}
    \begin{tabular}{l|c|cc|cc}\toprule
        \multirow{2}{*}{$g(x)$} & \multirow{2}{*}{$\log\ell_1$} &  \multicolumn{2}{c|}{Unnormalized} & \multicolumn{2}{c}{Normalized}\\
        & & $\ell_1^\text{STFT}$ & SDR & $\ell_1^\text{STFT}$ & SDR \\\midrule
Bypass-$g$                                  & \cellcolor{gray!99}{\color{white} -1.7431} & \cellcolor{gray!79}{\color{white} 0.2060} & \cellcolor{gray!99}{\color{white} 74.91} & \cellcolor{gray!79}{\color{white} 0.4702} & \cellcolor{gray!79}{\color{white} 65.24} \\
\midrule
HFF (\textsection\ref{sssec: hyperbolic})   & \cellcolor{gray!79}-3.5763 & \cellcolor{gray!99}{\color{white} 0.2905} & \cellcolor{gray!79}76.51 & \cellcolor{gray!99}{\color{white} 0.6474} & \cellcolor{gray!99}{\color{white} 61.76} \\
LFF (\textsection\ref{sssec: logarithmic})  & \cellcolor{gray!69}-3.7179 & \cellcolor{gray!79}{\color{white} 0.2101} & \cellcolor{gray!69}77.41 & \cellcolor{gray!69}0.4551 & \cellcolor{gray!79}{\color{white} 63.68} \\
PFF (\textsection\ref{sssec: power})        & \cellcolor{gray!59}-4.0731 & \cellcolor{gray!59}0.1020 & \cellcolor{gray!59}82.79 & \cellcolor{gray!59}0.2784 & \cellcolor{gray!59}67.18 \\
PQI (\textsection\ref{sssec: polynomial})   & \cellcolor{gray!29}-5.5426 & \cellcolor{gray!00}\textbf{0.0001} & \cellcolor{gray!39}112.56 & \cellcolor{gray!19}0.0004 & \cellcolor{gray!39}111.12 \\
\midrule
MLP(Identity)                               & \cellcolor{gray!49}-4.5614 & \cellcolor{gray!49}0.0040 & \cellcolor{gray!49}101.95 & \cellcolor{gray!49}0.0381 & \cellcolor{gray!49}85.51 \\
MLP($\tanh$)                                & \cellcolor{gray!19}-5.6983 & \cellcolor{gray!00}\textbf{0.0001} & \cellcolor{gray!29}128.42 & \cellcolor{gray!39}0.0006 & \cellcolor{gray!29}112.91 \\
MLP(ReLU)                                   & \cellcolor{gray!00}\textbf{-5.8888} & \cellcolor{gray!19}0.0003 & \cellcolor{gray!19}129.35 & \cellcolor{gray!29}0.0005 & \cellcolor{gray!19}114.85 \\
MLP($\sin$)                                 & \cellcolor{gray!39}-5.3744 & \cellcolor{gray!00}\textbf{0.0001} & \cellcolor{gray!00}\textbf{133.01} & \cellcolor{gray!00}\textbf{0.0003} & \cellcolor{gray!00}\textbf{116.88}
    \\\bottomrule\end{tabular}
\end{table}

\autoref{tab: results} provides a comparative analysis of objective scores, color-coded to depict performance from the worst (darkest cells) to the best (lightest cells) among $f^\dagger$ models.
Bold texts indicate the best scores and white texts indicate the less performant scores compared to Bypass-$g$.
Apart from the proposed MLP(sin) model, the table includes an ablation study on each MLP's nonlinearity functions (MLPs, HFF, LFF, and PFF), all trained using E2E.

The $\log\ell_1$ score reveals that MLP(ReLU) attains the superior performance.
This aligns with the findings depicted in \autoref{fig: id}, as the average $|\lambda_t-\hat{\lambda}_t|$ for MLP(ReLU) is lower, especially in small strain regions.
However, when subjected to evaluation using sinusoidal signals, MLP(sin) excels notably in both $\ell_1^{\text{STFT}}$ and SDR scores, demonstrating its prowess in both normalized and unnormalized scenarios.
This disparity in performance suggests that while MLP(ReLU) converges better, MLP(sin) better suppresses distortions, showcasing fewer harmonic artifacts.
This is analogous to why PQI, despite its inferior $\log\ell_1$ score, yields less harmonic than MLP (ReLU).

Notably, HFF and LFF, despite significantly improving the $\log\ell_1$ score, perform worse than the Bypass-$g$ case in some indices, indicating that incomplete compensation might exacerbate distortions.
PFF outperforms HFF or LFF, yet MLP(Identity), equipped solely with a Softplus as an activation function, surpasses PFF across all measures, reaffirming the approximation capability of MLPs.
However, PQI consistently outperforms MLP(Identity) across indices, closely trailing MLP(sin) with minor differences observed in $\ell_1^{\text{STFT}}$, yet displaying some shortcomings in SDR.

We also conduct a comparative analysis between the results obtained from the unnormalized stretch and the min-max normalized stretch signals.
The unnormalized stretch primarily consists of values within the range of $[0.96, 1.0]$ in the signal.
Given the limited range of stretch values, distinguishing between methods might pose some subtleties.
However, normalization proves to be instrumental in highlighting differences between the methods.
This emphasizes that MLP(sin) outperforms others, particularly among those equally ranked first in the unnormalized $\ell_1^{\text{STFT}}$ metric. 
These objective scores are further supported by an alignment with the actual samples.

\autoref{fig: spec} presents spectrograms of compensated outputs, showcasing harmonics aligned with results in \autoref{tab: results}.
Notably, MLP(sin) alongside PQI predominantly displays mere harmonics in its estimated spectrogram.
Although subtle, MLP(ReLU) manifests a discernible harmonic, while the remaining baselines distinctly present clear harmonics.
This again highlights the use of the periodic activation functions in MLPs.

\begin{figure}[t]
    \centering
    \begin{minipage}[b]{0.23\linewidth}
        \centering
        \centerline{\includegraphics[width=\linewidth]{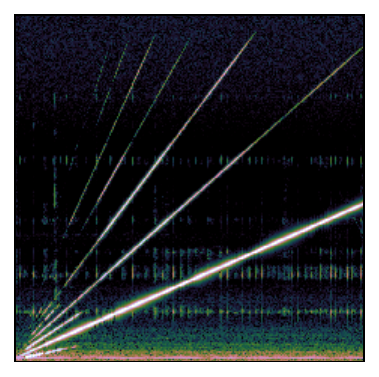}}
        \vspace{-2mm}
        \centerline{\footnotesize (a) Sinusoids}
    \end{minipage}
    \begin{minipage}[b]{0.23\linewidth}
        \centering
        \centerline{\includegraphics[width=\linewidth]{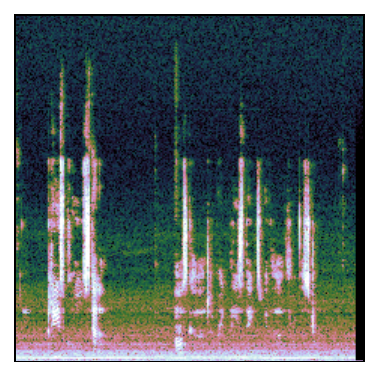}}
        \vspace{-2mm}
        \centerline{\footnotesize (b) Speech}
    \end{minipage}
    \begin{minipage}[b]{0.23\linewidth}
        \centering
        \centerline{\includegraphics[width=\linewidth]{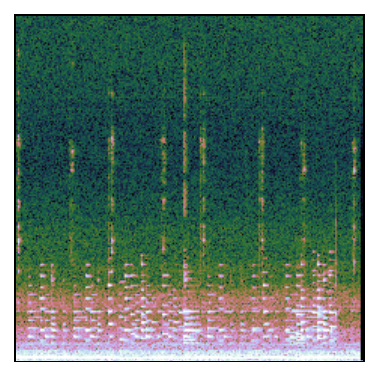}}
        \vspace{-2mm}
        \centerline{\footnotesize (c) Monophonic}
    \end{minipage}
    \begin{minipage}[b]{0.23\linewidth}
        \centering
        \centerline{\includegraphics[width=\linewidth]{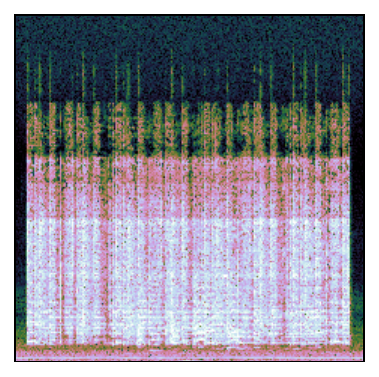}}
        \vspace{-2mm}
        \centerline{\footnotesize (d) Polyphonic}
    \end{minipage}
    
    \begin{minipage}[b]{0.23\linewidth}
        \centering
        \centerline{\includegraphics[width=\linewidth]{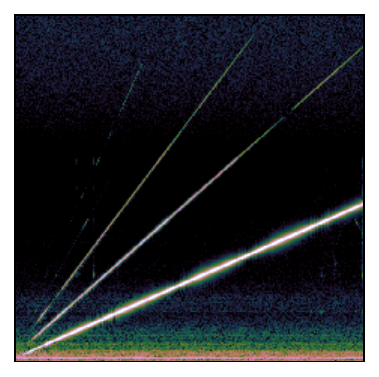}}
        \vspace{-2mm}
        \centerline{\footnotesize (e) Sinusoids}
    \end{minipage}
    \begin{minipage}[b]{0.23\linewidth}
        \centering
        \centerline{\includegraphics[width=\linewidth]{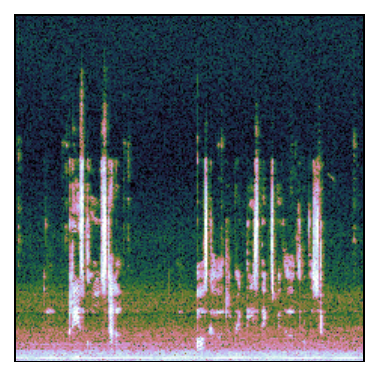}}
        \vspace{-2mm}
        \centerline{\footnotesize (f) Speech}
    \end{minipage}
    \begin{minipage}[b]{0.23\linewidth}
        \centering
        \centerline{\includegraphics[width=\linewidth]{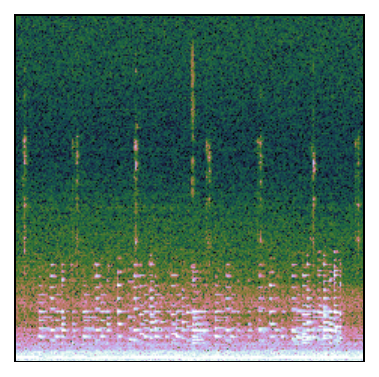}}
        \vspace{-2mm}
        \centerline{\footnotesize (g) Monophonic}
    \end{minipage}
    \begin{minipage}[b]{0.23\linewidth}
        \centering
        \centerline{\includegraphics[width=\linewidth]{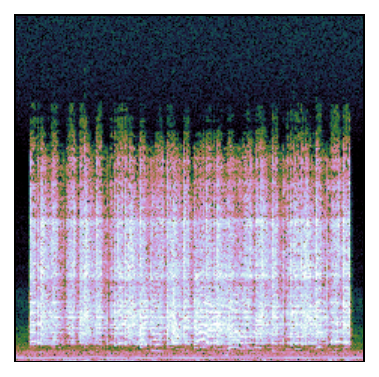}}
        \vspace{-2mm}
        \centerline{\footnotesize (h) Polyphonic}
    \end{minipage}
    \caption{\label{fig: measurement}{
        Spectrograms of the measured sound actuated using DEA.
        The signals without compensation are plotted on the top row.
        The signals compensated with MLP(sin) are plotted on the bottom row.
    }}
    \vspace{-6mm}
\end{figure}


To further validate the effectiveness of the compensation, we conduct an evaluation of MLP(sin) on real-world DEAs.
Given that our study involves simulations and compensations based on idealized DEAs, it is anticipated that the compensation results obtained from the manufactured DEAs might display discrepancies when compared to simulated outcomes.
For a comprehensive understanding of the equipment configuration, we direct readers to Appendix \ref{sec: appendix measurement setup} for detailed information.
Diverse signals encompassing sinusoidal sweep, speech, and musical inputs were pre-compensated and amplified before being applied to the DEA.

\autoref{fig: measurement} depicts spectrograms representing the measured signals.
Despite potential measurement errors and background noise, the spectrogram demonstrates a noteworthy reduction in harmonics following calibration compared to pre-calibration.
This observation holds even in cases where the original signal contains multiple frequency components, such as speech or music, revealing a significant improvement post-calibration.
This enhancement is particularly evident in a sine sweep input, but it becomes even more pronounced when dealing with signals containing a broader spectrum of frequencies.
For the sine sweep input, compensation using MLP(sin) yielded a result of 2.1\% for Total Harmonic Distortion (THD), marking a significant enhancement from the value of 14.89\% for the case without any compensation.

\section{Discussion and Limitation}
As demonstrated by the findings presented in the previous section, the proposed methodology employs numerical integration and neural networks to achieve efficient approximations while maintaining accuracy.
We develop a neural network to obtain the Runge-Kutta solution for the dielectric elastomer model more efficiently.
This model can be used to train another model that compensates for the nonlinearity of dielectric elastomers in an end-to-end framework, and we demonstrate its effectiveness.
In contrast to typical numerical integration methods that proceed through an internal loop until the solution converges, the trained neural network not only estimates the approximated solution in a single step but also enables INC through gradient backpropagation.

Despite these advances, there is potential room for improvements to be addressed.
Primarily, the proposed methodology is based on the assumption that DEA has a reasonably idealized form of deformation, and therefore may not fully capture nonlinearities in buckling or polarization. 
Also, the methodology has been validated exclusively for the properties outlined in \autoref{tab: params} and within a specified voltage range.
It is important to note that the values of these properties can fluctuate depending on the surrounding environment, including temperature and humidity.
This could potentially introduce inaccuracies in real-world measurement scenarios.

In the future, a number of methodologies could be employed to enhance the alignment between simulation and reality.
For instance, if temperature- and humidity-dependent property data is accessible, it may be feasible to utilize it as a training condition for the neural network.
Also, adaptively updating the neural network parameters to minimize the distortion from the measured microphone signal can also be a possible development, given that the neural network is sufficiently lightweight.

\section{Conclusion}
This study addresses the nonlinear deformation induced by dielectric actuation in pre-stressed ideal dielectric elastomers.
The paper establishes a comprehensive understanding by formulating a nonlinear ordinary differential equation based on the hyperelastic deformation, elucidating the intricate solution between voltage and stretch.
By utilizing numerical integration and neural networks, the solutions are efficiently approximated, without losing their accuracy.
The methods are evaluated through inverse nonlinearity compensation tasks showcasing the efficacy of the end-to-end trained MLPs.
The demonstrated effectiveness of these approximations, notably in minimizing harmonic distortions when utilized in acoustic actuation, underscores the significance of this research in enhancing the performance of such materials.


\appendices

\section{Details of the Evaluation}\label{sec: evaluation details}

\subsection{Choice of the forward deformation model}\label{ssec: forward deformation model}
The approximate solution was used because it is difficult to obtain the solution when applying a signal in addition to numerical stability to use RK45 as it is.
While it is possible to use a trained MLP model as the forward $f$, we do not see a significant performance improvement compared to the PQI, where the latter is more convenient in tuning its accuracy through adjusting $\tau_r$.

\subsection{Details on the inverse nonlinearity compensation function}\label{ssec: inc function}
As discussed in \autoref{sec: evaluation}, the inverse nonlinearity compensator of $f$ can be represented using an affine transformation of $f^{-1}$.
For evaluation, we utilize \autoref{eqn: g} with $\alpha=1000$, $a_1\approx-0.0341325$, and $a_0\approx 0.9999924$.
This particular choice of parameters is set to preserve the stretch at $V_{\text{min}}$ and $V_{\text{max}}$, \textit{i.e.}, $\lambda_{\text{min}}=f(V_{\text{min}})=f(\alpha g(V_{\text{min}}))$ and $\lambda_{\text{max}}=f(V_{\text{max}})=f(\alpha g(V_{\text{max}}))$.

\subsection{Training strategy}
Some may wonder about the efficacy of end-to-end training of $g_\theta$, as it is also possible to simulate the paired $(V,\lambda_t)$ for training $g_\theta$.
The performance of $g_\theta$s trained through this pseudo-code is already shown in \autoref{fig: id} (labeled MLP).
As the figure shows, no significant improvement was found through this training strategy, and the performances were typically worse than the models undergone E2E training.

%
%


\subsection{Computational efficiency}\label{ssec: computational efficiency}
We also compare the computation time between RK45, PQI, and MLP(sin).
\autoref{fig: time} summarizes the results.
To measure the computational time, we fix the hardware to either a single CPU thread (AMD Ryzen 5 3600 @2.2GHz) or a single GPU (RTX 2060).
The times are averaged over 1000 repeated measurements. 

Not surprisingly, MLP shows that its time does not vary with tolerance, while RK45 and PQI show a slowdown in computation as tolerance becomes smaller.
The number of samples used to measure the simulation time under $\tau_r$ is $10^3$.
Smaller values of tolerance usually increase the number of iterations required for the solution to converge, so it's not surprising that RK45 slows down.
However, it is interesting to see that the computation time of PQI increases with tolerance, even though no internal iterations are required.

Conversely, for compute time as a function of sample count, RK45 does not show a significant slowdown, while for PQI and RK54, the slowdown becomes apparent as soon as the number of samples exceeds $10^3$.
While it is not surprising that the total computation time increases as the number of samples increases, it is impressive to see that they are slower than the RK45 solver once the number of samples exceeds $10^4$.
We speculate that memory overhead may play a role in this slowdown.

\begin{figure}
    \centering
    \includegraphics[width=0.8\linewidth]{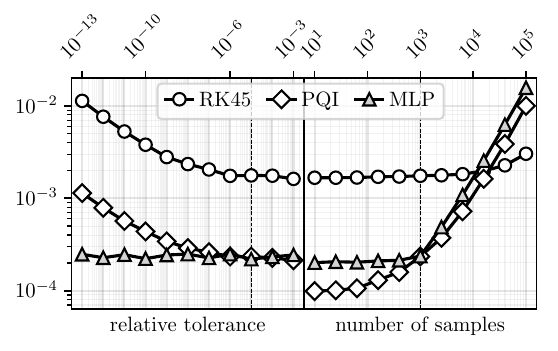}
    \vspace{-5mm}
    \caption{\label{fig: time}{
    computational time comparison
    }}
    \vspace{-5mm}
\end{figure}


\section{Baseline details}\label{sec: appendix baseline details}
In addition to PFFs using power functions, we also validate the compensation capabilities with trainable forms for other kinds of functions.
As two examples of such functions, this section provides some more details about LFF and HFF.

\begin{table}[t]
    \centering
    \caption{\label{tab: initialization} Initial parameters}
    \begin{tabular}{lrrrrr}\toprule
        Model & $\theta_{0}$ & $\theta_{1}$ & $\theta_{2}$ & $\theta_{3}$ & $\theta_{4}$ \\\midrule
    Baseline1 & $0.5$ & $3\times10^{3}$ & 0 & $1\times10^3$ & 0 \\
    Baseline2 & $0.3$ & $1\times10^{3}$ & 1 & $2\times10^3$ & 0 \\
    Baseline3 &     - & $1\times10^{4}$ & 0 & $9\times10^2$ & 0
    \\\bottomrule\end{tabular}
\end{table}

\subsection{Functions with Trainable Parameters}
As a different kind of compensation function from the power function, we consider logarithmic and hyperbolic families.
With a proper design of choice and initialization, such functions can also compensate for the deformation $f$ to some extent.
Exemplar identity plots $f\circ f^\dagger$ for each baseline $f^\dagger$ are plotted in \autoref{fig: id baseline}.
Based on the observations in \autoref{fig: id baseline}, we initialize the trainable parameters of each function as \autoref{tab: initialization} and fine-tune them to compensate better.

\begin{figure}[t]
    \begin{minipage}[b]{0.325\linewidth}
        \centering
        \centerline{\includegraphics[width=\linewidth]{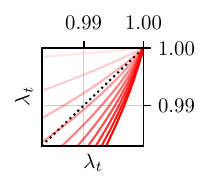}}
        \centerline{\footnotesize (a) PFFs}
    \end{minipage}
    \begin{minipage}[b]{0.325\linewidth}
        \centering
        \centerline{\includegraphics[width=\linewidth]{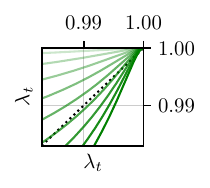}}
        \centerline{\footnotesize (b) LFFs}
    \end{minipage}
    \begin{minipage}[b]{0.325\linewidth}
        \centering
        \centerline{\includegraphics[width=\linewidth]{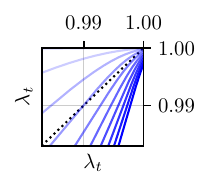}}
        \centerline{\footnotesize (c) HFFs}
    \end{minipage}
    \caption{\label{fig: id baseline}{Plot of $f\circ f^{\dagger}$.
    $\beta$ are swept from $10^2$ to $4\times10^3$ (curves for the larger $\beta$ are plotted in higher color intensity.)
    (a) $f^\dagger=10^3\sqrt{\beta_1(1-\lambda_t)}$, the first baseline as a square-root function.
    (b) $f^\dagger=\beta_2\log(10^3(1-\lambda_t^{0.3})+1)$, the second baseline as a logarithmic function.
    (c) $f^\dagger=10^4\tanh(\beta_3(1.002-\lambda_t)/45)$, the third baseline as a tanh function.
    }}
\end{figure}


\subsubsection{Logarithmic function}\label{sssec: logarithmic}
The logarithmic function fitting (LFF) is defined as follows.
\begin{equation}
    f^{(2)} = \theta_{1}^{(2)}\log\left(\theta_{2}^{(2)}\cdot\left(\theta_{3}^{(2)} -\lambda_t^{\theta_{0}^{(2)}}\right)+1\right)  + \theta_{4}^{(2)}
\end{equation}
Some might wonder why there also is a trainable power operator $\theta_0^{(2)}$.
The difference is marginal between fixing $\theta_0^{(2)}=1$ and updating it, but we use this power-ed version since it fits slightly better looking into \autoref{fig: id baseline} (b).

\subsubsection{Hyperbolic function}\label{sssec: hyperbolic}
We also study a hyperbolic family as the compensation function.
\begin{equation}
    f^{(3)} = \theta_{1}^{(3)} \tanh\left(\frac{\theta_{2}^{(3)}}{45}(1.002-\lambda_t)+\theta_{3}^{(3)}\right)  + \theta_{4}^{(3)}
\end{equation}
As $\tanh(x)=(e^x-e^{-x})/(e^x+e^{-x})$, this can also be viewed as a demonstration of an exponential family, but we denote this method as Hyperbolic Function Fitting (HFF).
Although we explored other families of trainable functions, such as logarithmic and hyperbolic functions, we observed no substantial improvement compared to the power function.

\section{Training Details}\label{sec: appendix training details}
In training the models, both $f_\theta$ and $g_\theta$ undergo training for 1M steps utilizing the Adam optimizer \cite{KingBa15}.
The learning rates are set to $\eta_f=10^{-6}$ and $\eta_g=10^{-5}$ for $f_\theta$ and $g_\theta$, respectively, and are subsequently reduced by a factor of $0.9$ every 20k steps.
This training configuration is identically applied to train MLPs and the baselines.

\begin{table}
    \centering
    \caption{Neural network architecture}
    \label{tab: network}
    \begin{tabular}{c|cc|l}\toprule
        Layer & Input dim. & Output dim. & Operations \\\midrule
        1 & 1   & 512 & Linear, sin, LayerNorm \\
        2 & 512 & 512 & Linear, sin, LayerNorm \\
        3 & 512 & 1   & Linear, Softplus
    \\\bottomrule\end{tabular}
\end{table}

\section{Measurement setup}\label{sec: appendix measurement setup}

\begin{figure}[ht]
    \begin{minipage}[b]{0.74\linewidth}
        \centering
        \centerline{\includegraphics[width=\linewidth]{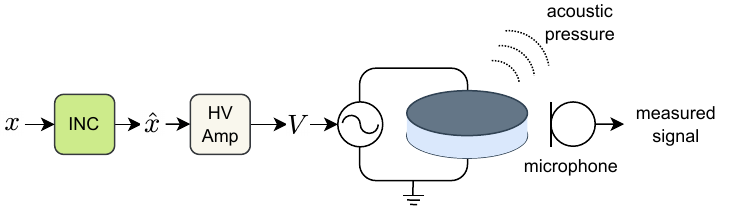}}
        \centerline{\footnotesize (a) Illustration of the measurement}
    \end{minipage}
    \begin{minipage}[b]{0.24\linewidth}
        \centering
        \centerline{\includegraphics[width=\linewidth]{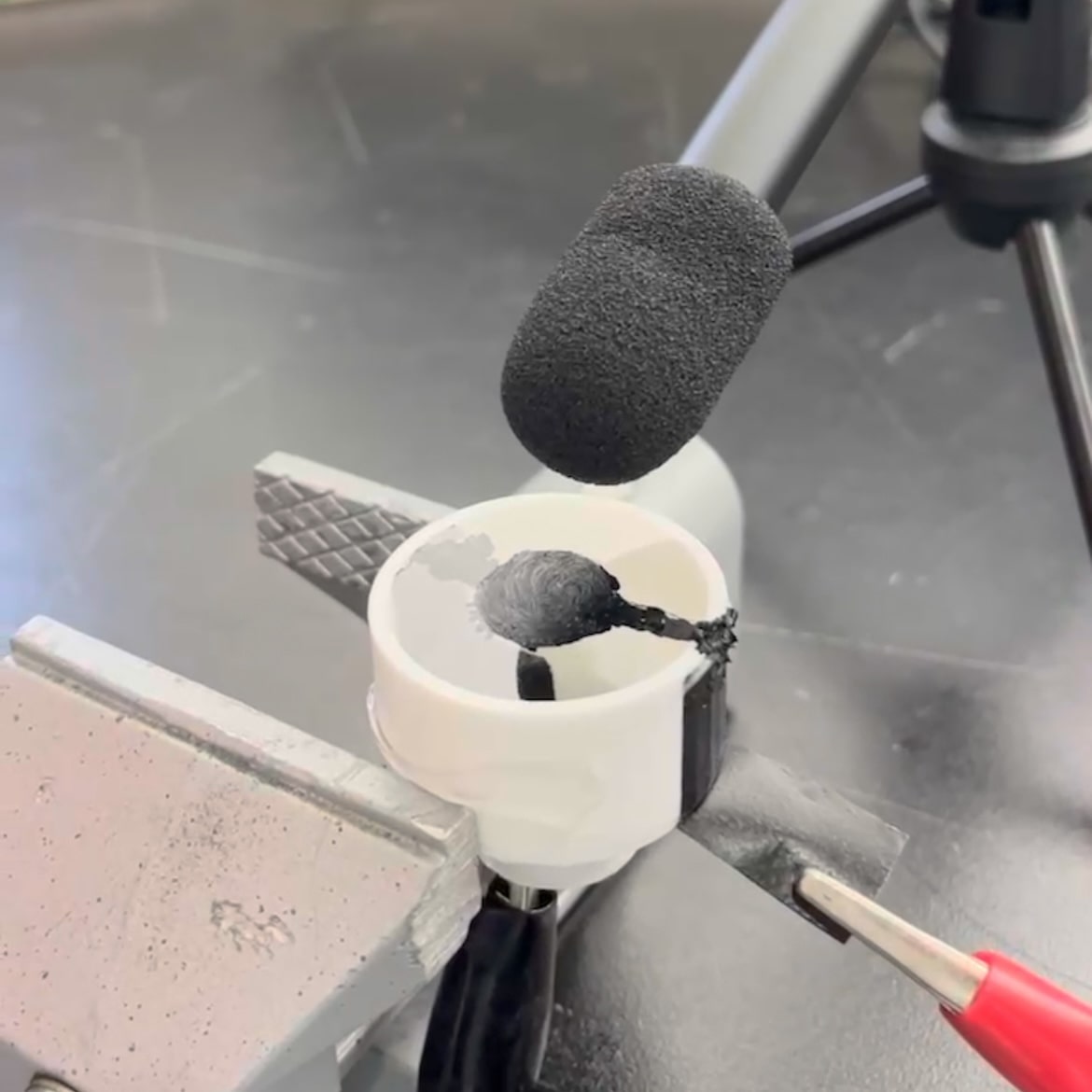}}
        \centerline{\footnotesize (b) DEA Sample}
    \end{minipage}
    \caption{DEA measurement equipment setup.
    }
    \label{fig: measurement-setup}
\end{figure}


We use \textsc{3M}\textsuperscript{\texttrademark} \textsc{VHB}\textsuperscript{\texttrademark} Tape 4910 as the elastomer of our DEA.
The elastomer is pre-stretched biaxially by a factor of 3.0, to be hung on a plastic frame.
We apply carbon grease on the elastomer to manufacture the dielectric elastomer.
So the VHB tape is sandwiched by the carbon grease electrode.
Regarding stability and Assumption \ref{assm: no buckling}, the electrodes are applied to only a portion of the center of the elastomer.
We amplify the voltage signal by a factor of 1000 to actuate the DEA.
The actuated sound is picked up using a Behringer\texttrademark\ ECM8000 microphone.
\autoref{fig: measurement-setup} shows a photo of part of the experimental setup.

\bibliographystyle{IEEEtran}
\bibliography{ref}

\begin{IEEEbiography}[{\includegraphics[width=1in,height=1.25in,clip,keepaspectratio]{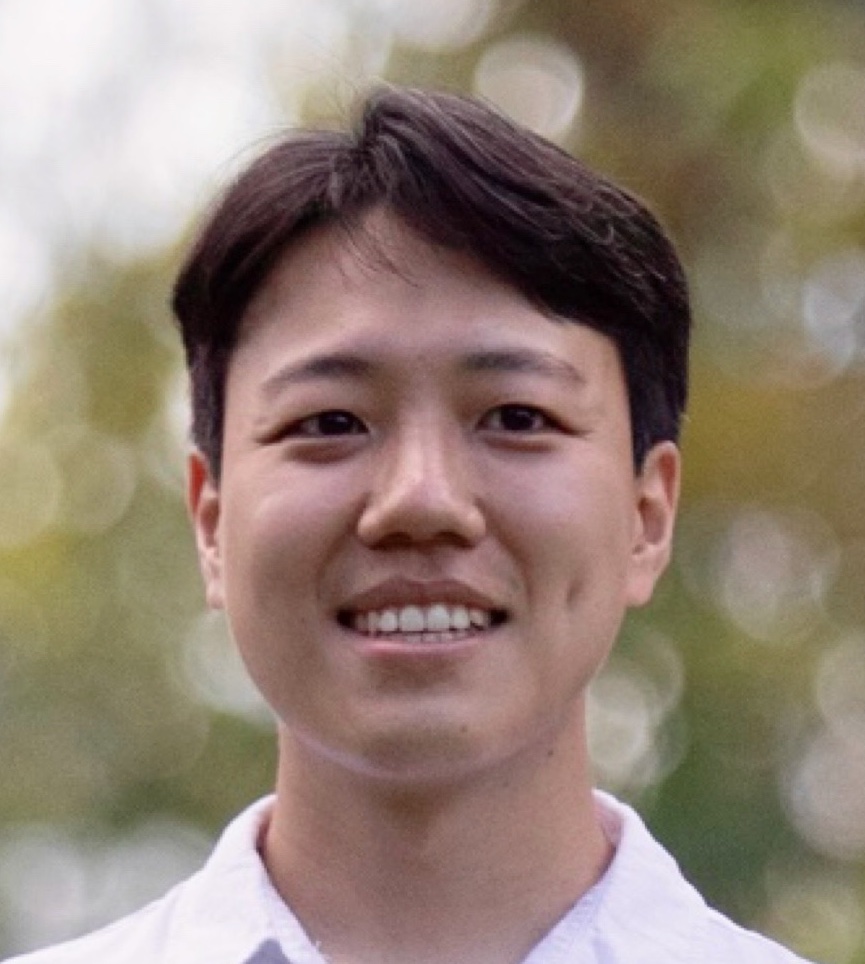}}]{Jin Woo Lee}
is a Ph.D. candidate at the Music and Audio Research Group
of Department of Intelligence and Information
at Seoul National University in South Korea.
He received the B.Sc. degree in Mechanical Engineering
from the Pohang University of Science and Technology,
Pohang, South Korea, in 2019. 
His research interest is in numerical sound synthesis,
physical modeling, and machine learning.
\end{IEEEbiography}

\begin{IEEEbiography}[{\includegraphics[width=1in,height=1.25in,clip,keepaspectratio]{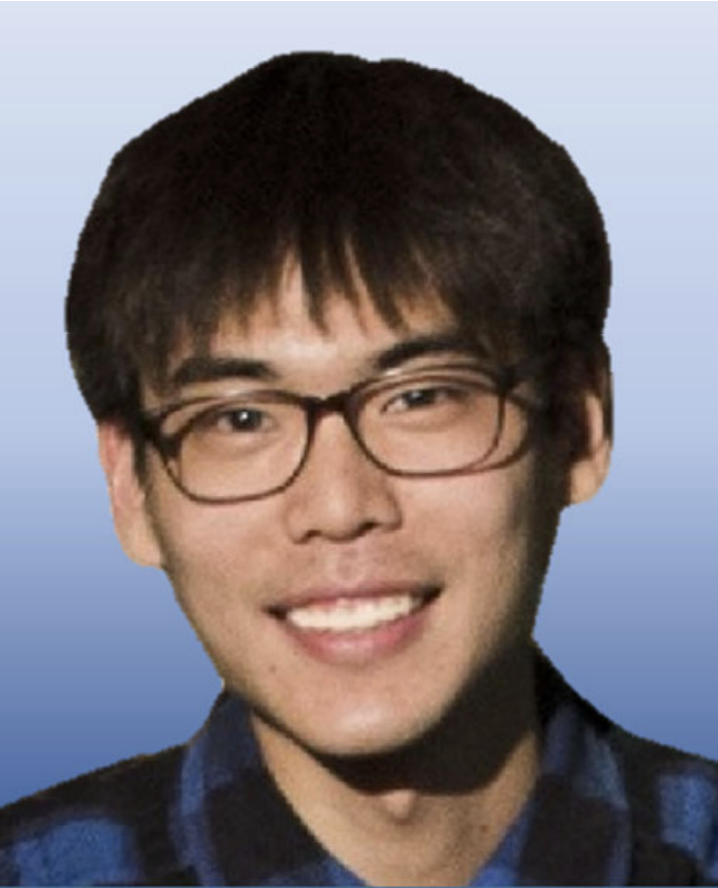}}]{Gwang Seok An}
is a Ph.D. candidate at the Music and Audio Research Group
in the Graduate School of Convergence Science and Technology at
Seoul National University in South Korea.
He received his B.S. degree in Electronic Engineering from
the Myongji University in 2012.
His research interest is in digital audio signal processing
for embedded systems.
\end{IEEEbiography}

\begin{IEEEbiography}[{\includegraphics[width=1in,height=1.25in,clip,keepaspectratio]{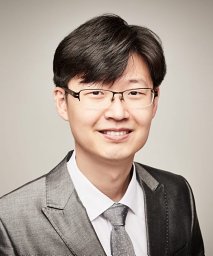}}]{Jeong-Yun Sun}
is a Professor in Materials Science and Engineering
at Seoul National University in South Korea.
He has been
a Research Associate in Material Science and Mechanical Engineering
and a Post Doctoral Fellow in School of Engineering and Applied Science
at Harvard University, MA, USA. 
He received the Ph.D. degree
in Materials Science and Engineering at Seoul National University.
He is currently with the Research Institute of Advanced Materials (RIAM) and Department of Materials Science and Engineering at Seoul National University.
His research interests include
ionics, biomedical engineering, and soft materials.
\end{IEEEbiography}

\begin{IEEEbiography}[{\includegraphics[width=1in,height=1.25in,clip,keepaspectratio]{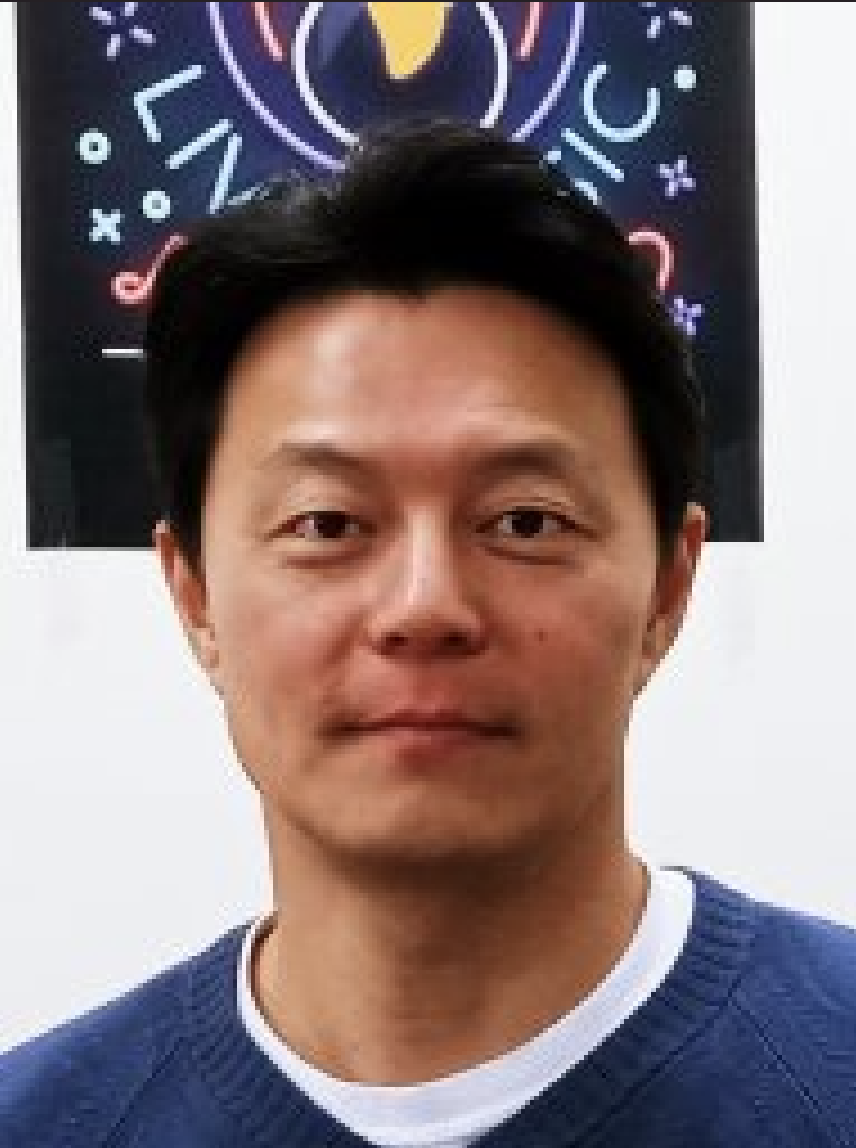}}]{Kyogu Lee}
is a Professor in Department of Intelligence and Information
at Seoul National University in South Korea and is a director of the
Music and Audio Research Group at Seoul National University.
He received the Ph.D. degree in Computer-based Music Theory and
Acoustics from Stanford University, CA, USA. 
He is currently with the Department of Intelligence and Information, Interdisciplinary Program in Artificial Intelligence (IPAI), and Artificial Intelligence Institute (AII), at Seoul National University.
His research interests include signal processing and machine learning
techniques applied to music, audio, and hearing.
\end{IEEEbiography}

\EOD

\end{document}